\documentclass[twocolumn,prl,amssymb]{revtex4}
\def\th{^{\mbox{\scriptsize th}}}

\def\id{\mathbb{I}}
\def\vs{\vspace{1.5mm}}

\newcommand{\ket}[1]{\mbox{$| #1 \rangle$}}
\newcommand{\bra}[1]{\mbox{$\langle #1 |$}}
\newcommand{\proj}[1]{\ket{#1}\!\bra{#1}}
\def\tr{\mbox{tr}}

\def\h{\mathcal{H}}

\def\ze{{\bf 0}}
\def\A{A_n(a|x)}

\begin{document}

\title{Asymptotic violation of Bell inequalities and distillability}
\author{Llu\'{\i}s Masanes}
\affiliation{School of Mathematics, University of Bristol, Bristol
BS8 1TW, U.K.}
\date{\today}

\begin{abstract}
A multipartite quantum state violates a Bell inequality
asymptotically if, after jointly processing by general local
operations an arbitrarily large number of copies of it, the result
violates the inequality. In the bipartite case we show that
asymptotic violation of the CHSH-inequality is equivalent to
distillability. Hence, bound entangled states do not violate it. In
the multipartite case we consider the complete set of
full-correlation Bell inequalities with two dichotomic observables
per site, also called WWZB-inequalities. We show that asymptotic
violation of any of these inequalities by a multipartite state
implies that pure-state entanglement can be distilled from it,
although the corresponding distillation protocol may require that
some of the parties join into several groups. We also obtain the
extreme points of the set of distributions generated by measuring
$N$ quantum systems with two dichotomic observables per site. It is
shown that when considering the violation of any Bell inequality
after preprocessing, either deterministic LOCC or stochastic local
operations (without communication) is enough.
\end{abstract}
\maketitle

In 1964 Bell ruled out the possibility that a local classical
theory could give the same predictions than Quantum Mechanics
\cite{Bell}. Bell's Theorem states that the probabilities for the
outcomes obtained when suitably measuring some quantum states
cannot be generated from classical correlations.
This turns out to be a general feature for entangled pure states, as
it was proven in \cite{nose} that, a multipartite pure states is
entangled (not product) if, and only if, it is not simulable by a
local variable model (LVM). The situation is more complex for mixed
states. There are mixed states that, though being entangled,
whenever a single copy of the state is measured there is always a
LVM giving the same predictions \cite{B,W}. But some of these states
violate Bell inequalities if, prior to the measurement, the state is
preprocessed \cite{G}. The most general preprocessing consists of
stochastic local operations and classical communication (SLOCC),
that is, LOCC protocols that fail with some probability. In this
paper it is shown that when considering violation of Bell
inequalities, both: preprocessing by stochastic local operations
without communication (SLO), and, preprocessing by (deterministic)
LOCC, are completely general \cite{deterministic}. It is important
to remark that, in order to build a consistent picture, the
preprocessing (in particular the classical communication) has to be
made before the parties choose the experimental settings for the
Bell test.

\vs

Another way of making manifest the nonlocality ``hidden" in a mixed
state is by jointly measuring more than one copy of it. In \cite{P}
it is shown that some states having a LVM for the single copy
scenario, violate Bell inequalities when jointly measuring more than
one copy. Then, merging these two ideas, a strong test for detecting
the nonlocality ``hidden'' in a state is to measure the result of
jointly processing by SLO an arbitrarily large number of copies of
the state (as mentioned above, SLO is completely general). If the
resulting probabilities violate a Bell inequality, we say that the
original state violates this inequality asymptotically. Clearly,
this kind of test is more general than above mentioned ones. That
is, the set of states that violate a Bell inequality asymptotically
includes the states that violate it straightaway. Later, we discuss
how general the notion of asymptotic nonlocality is, and
relate it to distillability. \vs

A bipartite state is said to be distillable if, from an
arbitrarily large number of copies of it, some pure-state
entanglement can be extracted by LOCC \cite{BDSW}. For states of
more than two parties different notions of distillability can be
considered. For instance, there are $N$-partite states which are
undistillable when the $N$ parties remain separated, but if some
of the parties join together, pure-state entanglement among the
different groups of parties can be obtained \cite{DC}. Here, we
also say that such states are distillable. \vs

Distillability and violation of Bell inequalities are two
manifestations of entanglement. On one hand, distillability is
related to the usefulness in quantum information processing tasks,
due to the fact that most of them require pure-state entanglement
as a principal ingredient. On the other hand, Bell violation
expresses the fact that a state cannot be simulated by classical
correlations. This seems to be also a requirement if we want that
a quantum information task gives an advantage over its classical
counterparts. In particular, violation of Bell inequalities is a
necessary and sufficient condition for the usefulness of a quantum
state in communication complexity tasks \cite{BZPZ}. In this
paper, the link between these two concepts is analyzed. In
particular, it is shown that asymptotic violation of the
CHSH-inequality \cite{CHSH} is equivalent to distillability. \vs

One of the most remarkable results in the field of Bell
inequalities is the complete characterization of all $N$-partite
inequalities with full-correlation functions of two dichotomic
observables per site \cite{WW,ZB}, here called WWZB-inequalities.
A relation between distillability and violation of
WWZB-inequalities for $N$-qubit systems was presented in
\cite{A,ASW,ASW2}. They showed that the violation of a
WWZB-inequality by a multiqubit state implies that pure-state
entanglement can be distilled from it, although the corresponding
distillation protocol may require that some of the parties join
into several groups. In this paper we generalize their results to
$N$-partite systems with arbitrary local Hilbert spaces (instead
of qubits). We also generalize these results to the asymptotic
scenario. In particular, given an arbitrary $N$-partite state,
asymptotic violation of a WWZB-inequality implies distillability
in the same sense as mentioned above. \vs

Let us specify the scenario and the notation. Consider $N$
separated parties, denoted by $n=1,\ldots N$, each having a
physical system which can be measured with one among $M$
observables with $K$ outcomes each. The $n\th$ party observables
and outcomes are respectively denoted by $x_n\in \{1,\ldots M\}$
and $a_n\in \{1,\ldots K\}$. All the experimental information is
in the joint probability distribution for the outcomes
conditioned on the chosen observables $P(a_1\ldots a_N|x_1\ldots
x_N)$. Distributions that correspond to LVM are the ones that can
be written as
\begin{equation}\label{CC}
    P(a_1\ldots a_N|x_1\ldots x_N) =
    \sum_\lambda p(\lambda) \prod_{n=1}^N P_n(a_n|x_n \lambda)
\end{equation}
see \cite{WWrev}. Fixed $N,M,K$ to some finite values, the set of
distributions $P$ that can be written as (\ref{CC}) is a convex
polytope, which can be characterized by a finite set of linear
inequalities \cite{WWrev}. Each of these inequalities, denoted
$\beta$, is characterized by its coefficients
\begin{eqnarray}\label{BI}
    \beta[P]=
    \sum_{a_1, x_1}\cdots\sum_{a_N, x_N}
    \beta(a_1\ldots a_N|x_1\ldots x_N) \\
    \times \nonumber
    P(a_1\ldots a_N|x_1\ldots x_N) \geq 0\ .
\end{eqnarray}
By definition, all distributions of the form (\ref{CC}) satisfy
(\ref{BI}). If one inequality $\beta$ is violated by some
probability distributions, we say that $\beta$ is a Bell inequality.
\vs

Let us characterize the set of distributions that can be generated
within quantum theory. Suppose the $n\th$ party has a system with
Hilbert space $\h_n$, which is measured with the $M$ generalized
measurements $\{\A:\ a=1,\ldots K\}$ for $x=1,\ldots M$. These
POVMs satisfy $\A\geq 0$ for $a=1,\ldots K$ and $\sum_{a=1}^K \A
=\id_n$, for $x=1,\ldots M$ and $n=1,\ldots N$, where $\id_n$ is
the identity matrix acting on $\h_n$.
The quantum distributions are the ones that can be written as
\begin{equation}\label{QC}
    P(a_1\ldots a_N|x_1\ldots x_N) =
    \tr\!\left[\rho\, \bigotimes_{n=1}^N A_n(a_n|x_n)
    \right]\ ,
\end{equation}
where $\rho$ is a positive semidefinite matrix acting on
$\h=\bigotimes_{n=1}^N \h_n$ with $\tr\rho=1$. It is shown in
\cite{WW} that the sets of quantum distributions (\ref{QC}) are
convex, but little is known about them. Here we obtain a
characterization of all the extreme points for the case $K=M=2$ and
arbitrary $N$. \vs

{\bf Lemma.} Let $A_1, A_2, B_1, B_2$ be four projectors acting on
a Hilbert space $\h$ such that $A_1 + A_2 =\id$ and $B_1 + B_2
=\id$. There exists an orthonormal basis in $\h$ where the four
projectors $A_1, A_2, B_1, B_2$ are simultaneously block-diagonal,
in blocks of size $1\times 1$ or $2\times 2$.\vs

{\em proof.} The three positive operators $B_1$, $(B_1 A_1 B_1)$,
$(B_1 A_2 B_1)$ can be simultaneously diagonalized, because their
ranges are contained in the subspace where $B_1$ acts like the
identity and $B_1 A_1 B_1 + B_1 A_2 B_1 = B_1$. Let $\ket{v}$ be one
of their simultaneous eigenvectors which satisfies $B_2 \ket{v}=
\ze$. Because $A_1+A_2 =\id$, it cannot be the case that $A_1
\ket{v} = A_2 \ket{v} = \ze$. If $A_1 \ket{v}=\ze$ then $A_2
\ket{v}= B_1 \ket{v}= \ket{v}$ and the span of $\ket{v}$ (denoted
$E_v$) corresponds to a $1\!\times\! 1$ diagonal block in which
$A_1, A_2, B_1, B_2$ have eigenvalues $0,1,1,0$, respectively. The
case $A_2 \ket{v} =\ze$ is similar. Consider the case where $A_1
\ket{v}\neq \ze$ and $A_2 \ket{v} \neq\ze$. Define the orthogonal
vectors $\ket{a_1}=A_1\ket{v}$, $\ket{a_2}=A_2\ket{v}$, and the
two-dimensional subspace $E_v= \{\alpha_1\ket{a_1} +
\alpha_2\ket{a_2}:\, \forall \alpha_1, \alpha_2 \in \mathbb{C}\}$.
The fact $\ket{v}= \ket{a_1} + \ket{a_2}$ implies $\ket{v} \in E_v$.
Because $B_1 \ket{a_1} \!\propto\! \ket{v}$ and $B_1 \ket{a_2}
\propto \ket{v}$, there exists a vector $\ket{w}\in E_v$ such that
$B_1\ket{w}= \ze$ and $B_2\ket{w}= \ket{w}$. Summarizing, the
vectors $\ket{a_1}, \ket{a_2} \in E_v$ are simultaneous eigenvectors
of $A_1, A_2$, and the vectors $\ket{w}, \ket{v} \in E_v$ are
simultaneous eigenvectors of $B_1,B_2$. Therefore, the subspace
$E_v$ corresponds to a $2\!\times\! 2$ simultaneous diagonal block
for $A_1, A_2, B_1, B_2$. The same can be done with the rest of
simultaneous eigenvectors $\ket{v}$
as defined above. 
And analogously, for the simultaneous eigenvectors of $B_2$, $(B_2
A_1 B_2)$, $(B_2 A_2 B_2)$ which are orthogonal to the vectors
$\ket{w}$ that have appeared in the previous steps. At the end, the
direct sum of the subspaces $E_1, E_2 \ldots$ is $\h$, each subspace
$E_i$ of dimension two contains two eigenvectors of each operator
$A_1, A_2, B_1, B_2$. \hfill $\Box$ \vs

{\bf Result 1.} In the case $K=M=2$, all quantum extreme points
(\ref{QC}) are achievable by measuring $N$-qubit pure states with
projective observables. \vs

{\em Proof.} It is easy to see that all two-outcome POVMs are
mixtures of two-outcome projective measurements. Hence, the
distribution (\ref{QC}) can be written as a mixture of distributions
where the operators $\A$ are projectors. According to the lemma, the
four operators $\A$ for $a,x=1,2$ can be simultaneously
block-diagonalized, in blocks of size $2\times 2$, at most. Denote
by $\{E_1^n, E_2^n\ldots \}$ the projectors onto the subspaces
corresponding to these diagonal blocks, for the $n\th$ party. If the
$n\th$ party performs the measurement $\{E_i^n\}_i$ before measuring
$\{A(1|x), A(2|x)\}$, the result does not change. But, after
performing $\{E_i^n\}_i$ the local system is contained in a
two-dimensional subspace. Applying this to all parties, the
distribution (\ref{QC}) becomes a mixture of distributions generated
by measuring $N$-qubit systems with projective observables. To
finish, recall that $\rho$ can be expressed as a mixture of pure
states. \hfill $\Box$ \vs

{\bf Result 2.} If an $N$-partite state $\rho$ violates the Bell
inequality $\beta$ (for $M=K=2$), then $\rho$ can be transformed by
SLO into an $N$-qubit state $\tilde{\rho}$ that violates $\beta$ by
an equal or larger amount.\vs

{\em Proof.} If $\rho$ violates $\beta$ then it does so with
projective observables, because these are more extremal. Following
the argument of the previous proof, the correlations obtained from
$\rho$, say (\ref{QC}), do not change if the $n\th$ party performs
the measurement $\{E_i^n\}_i$ before measuring $\A$, for all $n$.
The final distribution (\ref{QC}) becomes a mixture of
distributions generated by the family of two-qubit states
$(E^1_{i_1}\otimes \cdots \otimes E^N_{i_N}) \rho\,
(E^1_{i_1}\otimes\cdots\otimes E^N_{i_N})$. By convexity, at least
one of these states violates $\beta$ by at least the same amount.
\hfill $\Box$ \vs

As mentioned above, the set of states that violate a Bell inequality
after a general preprocessing (SLOCC) is strictly larger than the
set of states that violate Bell straightaway \cite{G}. It could also
be the case that, the set of states that violate Bell after SLOCC is
larger than the set of states that do so after SLO, or
(deterministic) LOCC; due to the fact that these two classes of
operations are more restricted. Remarkably, the three sets are
equally powerful. Recall that some authors also consider Bell
inequalities which are not facets of the LVM-polytope. This result
does not hold for them. \vs

{\bf Result 3.} If a state violates a Bell-inequality $\beta$
after SLOCC, then it also violates $\beta$ after SLO or LOCC.\vs

{\em Proof.} To see that SLO is enough, recall that any SLOCC
protocol can be pictured as a tree where different branches
correspond to different strings of exchanged messages, and the
result of at least one branch must violate $\beta$. The local
operations corresponding to this branch constitute a SLO protocol
which also achieves this goal. To show that LOCC is enough, suppose
that $\Omega$ is a SLOCC protocol that can be applied to the state
$\rho$ with probability $\pi= \tr[\Omega(\rho)]$, and the state
$\Omega(\rho)$ violates the Bell inequality $\beta$ when measured
with the observables $\{A_n(a_n|x_n)\}$. Suppose that $P_0$ is a LVM
distribution (\ref{CC}) that saturates the inequality, $\beta[P_0
]=0$. Given $P_0$ one can always get a separable state $\rho_0$ and
a set of observables $\{B_n(a_n|x_n)\}$ such that
\begin{equation}
    P_0(a_1\ldots a_N|x_1\ldots x_N) =
    \tr\!\left[\rho_0 \bigotimes_{n=1}^N B_n(a_n|x_n)
    \right]. \nonumber
\end{equation}
Define a LOCC protocol $\tilde{\Omega}$ in the following way. The
parties perform the protocol $\Omega$ to $\rho$, if it succeeds
each party prepares an ancillary system in the state $\ket{0}$, if
it fails each party prepares an ancillary system in the state
$\ket{1}$ and all together prepare the separable state $\rho_0$
(by LOCC), and finally, all parties forget the state of the
ancillary systems. The state $\rho$ after this protocol is
\begin{equation}
    \tilde{\Omega}(\rho)= \pi\, \Omega(\rho)\otimes
    \proj{0}^{\otimes N} + (1-\pi) \rho_0 \otimes\proj{1}^{\otimes N}\ .
\end{equation}
Clearly, if this state is measured with the observables
$\{A_n(a_n|x_n)\otimes\proj{0} + B_n(a_n|x_n)\otimes\proj{1}\}$
one obtains a strict violation of $\beta$, as we wanted to prove.
\hfill $\Box$ \vs

Let us move to the asymptotic scenario. In the rest of the paper we
only consider the case where each party has two dichotomic
observables ($M=K=2$). From Result 2, straight conclusions can be
obtained in the bipartite case.\vs

{\bf Result 4.} A bipartite state $\rho$ is distillable if, and
only if, there exist a positive integer $m$ and a SLO map
$\Omega$ such that $\Omega[\rho^{\otimes m}]$ violates CHSH.\vs

{\em Proof.} By definition, if $\rho$ is distillable there exist
an integer $m$ and a SLO map $\Omega$ such that the state
$\Omega[\rho^{\otimes m}]$ is close enough to a singlet for
violating CHSH. Conversely, if there exists an integer $m$ and a
SLO map $\Omega$ such that $\Omega[\rho^{\otimes m}]$ violates
CHSH, by Result 2, the state $\Omega[\rho^{\otimes m}]$ can be
transformed by SLO into a two-qubit state $\tilde{\rho}$ which
violates CHSH. Clearly, $\tilde{\rho}$ is entangled and thus
distillable \cite{HHH}, therefore $\rho$ is distillable too.
\hfill $\Box$ \vs

Let us generalize this result to the multipartite case. When the
observables are dichotomic ($K=2$), one can reduce the amount of
experimental data by considering full-correlation functions
\begin{eqnarray}\label{cf}
    && C(x_1\ldots x_N)=\\ \nonumber
    && \sum_{a_1=1}^2 \ldots \sum_{a_N =1}^2
    (-1)^{\sum_{n=1}^N a_n}
    P(a_1\ldots a_N|x_1\ldots x_N)\ .
\end{eqnarray}
That is, for each experimental setting $(x_1\ldots x_N)$ all the
information is summarized in the single number $C(x_1\ldots x_N)$,
instead of the $2^N$ numbers $P(a_1\ldots a_N|x_1\ldots x_N)$. In
the case where each party has two dichotomic observables
($M=K=2$), the necessary and sufficient conditions for a given
sample of correlators $\{C(x_1\ldots x_N)\}$ to be obtainable by a
LVM are the WWZB-inequalities \cite{WW,ZB}. A link between
violation of WWZB-inequalities and distillability for $N$-qubit
systems was obtained in \cite{A,ASW,ASW2}. They proved that, if an
$N$-qubit state $\rho$ violates a WWZB-inequality $\beta$ by an
amount $\beta[\rho]$ such that
\begin{equation}\label{}
    1<2^{\frac{N-G-1}{2}}<\beta[\rho] \leq 2^{\frac{N-G}{2}}\ ,
\end{equation}
then pure-state entanglement can be extracted from $\rho$ when the
parties join into groups of at most $G$ people. That is, the
larger the violation is, the smaller the size of the groups $G$
has to be in order to distill. \vs

Result 2 straightforwardly allows for generalizing this results to
any $N$-partite state $\rho$, not necessarily composed of qubits.
Also, following the same reasoning as in Result 4, one can
generalize it to hold in the asymptotic scenario. \vs

{\bf Result 5.} Consider an $N$-partite state $\rho$, an integer
$m$ and a SLO map $\Omega$ such that the WWZB-inequality $\beta$
is asymptotically violated by the amount
$\beta[\Omega(\rho^{\otimes m})]$ in the range
\begin{equation}\label{}
    1<2^{\frac{N-G-1}{2}}<\beta[\Omega(\rho^{\otimes m})]
    \leq 2^{\frac{N-G}{2}}\ .
\end{equation}
Then, pure-state entanglement can be extracted from $\rho$ when
the parties join into groups of at most $G$ people.\vs

{\bf Discussion.} When considering violations of Bell inequalities,
the asymptotic scenario is more general than that with single-copy
preprocessing \cite{G}, and joint measurements on few copies of the
state \cite{P}. However, there is another scenario which in
principle is not included in the asymptotic one, namely, when
sequences of more than one measurement in each site are applied to
the same system \cite{Po2}. Interestingly, the only known strategy
to apply sequences of measurements \cite{Po2} in order to violate
Bell inequalities is equivalent to SLO-preprocessing of the state
\cite{P2}. Then, the asymptotic scenario may be completely general
when considering Bell violation with a single species of state.  \vs

We have seen that in the bipartite case, asymptotic nonlocality is
equivalent to distillability, if we impose that the Bell
experiment is made with two dichotomic observables per site
($K=M=2$). As byproduct we obtain a new result: bound entangled
states \cite{be} do not violate CHSH, even asymptotically. It
would be very interesting to know whether the above equivalence
holds without any restriction on $K$ and $M$. Unfortunately, the
techniques used in the lemma cannot be directly extended to larger
values of $K$ or $M$.
In the multipartite case, we have generalized the results obtained
in \cite{A,ASW,ASW2} to arbitrary states (not only multi-qubit
states), and also for the asymptotic scenario.\vs

There is another scenario which is more general than the one of
asymptotic nonlocality. Namely, when a bipartite state cannot be
simulated by classical correlations in all possible situations
\cite{ANS}. These include situations when the bipartite state
$\rho$ is jointly processed with a different state $\omega$. In
the case where $\omega$ is equal to $\rho^{\otimes n}$ we recover
the asymptotic scenario. It is proven in \cite{ANS} that a
bipartite state is simulable by classical correlations in all
possible situations if, and only if, it is separable. Merging this
equivalence with the one of Result 4 an appealing picture for
bipartite states emerges:
\begin{eqnarray*}
  \mbox{entangled}  &\Longleftrightarrow&
  \mbox{nonsimulable in general} \\
  \mbox{distillable} &\Longleftrightarrow&
  \mbox{nonsimulable in the asymp. scen.}
\end{eqnarray*}
Unfortunately, the second equivalence is only proved for the case
$K=M=2$. But as argued above, it could hold in general. \vs

\vs We have also studied the set of probability distributions that
arise when measuring $N$ quantum systems with two dichotomic
observables per site. We have shown that the extreme points of this
convex set are obtainable by measuring $N$-qubit pure states with
projective observables. This answers the question posed in Problem
26A of \cite{KW}, for the case $M=K=2$. It would be very interesting
to generalize this result by showing that, the minimal dimension of
the local Hilbert space sufficient for generating all extreme
quantum correlations in the setting $(N,M,K)$ is $K$. Unfortunately,
the technique used in the lemma is not directly applicable for $M,K
>2$. These bounds on the dimensionality of the local Hilbert space
allow for using known algorithms \cite{algol} in order to decide
whether some correlations are predictable by quantum theory or not.
This could be used to falsate quantum mechanics without assuming any
model for the experiment, as Bell inequalities do so for classical
physics.

{\bf Acknowledgements.} The author is thankful to Nick S. Jones for
useful comments. This work has been supported by the U.K. EPSRC's
``IRC QIP''.


\begin{thebibliography}{99}
\bibitem{deterministic} A deterministic operation is one that
succeeds with probability one (trace-preserving).
\end{thebibliography}
\end{document}